\documentclass[11pt]{article}

\usepackage{bm}
\usepackage{amsfonts,amsmath}
\usepackage{amsmath}
\usepackage{amssymb}
\usepackage{amsthm}
\newtheorem{pro}{Proposition}

\newtheorem{theo}{Theorem}
\date{}
\usepackage[dvips]{color}
\usepackage{color}
\usepackage{graphicx}
\usepackage{exscale,relsize}

\title{\textbf{The Einstein-Hilbert action of the space of holomorphic maps from $S^2$ to $\mathbb{C}P^k$}}
\author{L.S. Alqahtani \\
\small \textit{Department of Pure Mathematics, University of Leeds}\\
\small \textit{Leeds LS2 9JT, UK}\\
\small \texttt{ mmlsa@leeds.ac.uk}
}

\begin{document}

\maketitle\begin{abstract}
Let $\mathcal{H}_{n,k}(\Sigma )$ be the space of degree $n\geq 1$  holomorphic maps from a compact Riemann surface $\Sigma $ to $\mathbb{C}P^k$. In the case $\Sigma=S^2$ and $n=1$, the $L^2$ metric on $\mathcal{H}_{1,k}(S^2 )$ was computed exactly by Speight. In this paper, the Ricci curvature tensor and the scalar curvature on $\mathcal{H}_{1,k}(S^2 )$ are determined explicitly for $k\geq 2$. An exact direct computation of the Einstein-Hilbert action with respect to the $L^2$ metric on $\mathcal{H}_{1,k}(S^2 )$ is made and shown to coincide with a formula conjectured by Baptista.
\end{abstract}

\section{Introduction}

Let $\Sigma $ be a compact Riemann surface equipped with a Riemannian metric  $g$ and let $h$ be the Fubini-Study metric on  $\mathbb{C}P^k$. Let $\phi $ be a holomorphic map from $\Sigma $  to $\mathbb{C}P^k$ of  degree $n\geq 1$ defined as

\begin{equation}
n=\int_{\Sigma } \phi ^* \omega _0,
\end{equation}

\noindent where $\omega _0$ is the normalized K\"{a}hler form with respect to $h$. Consider the   space of degree $n$  holomorphic maps $\Sigma \rightarrow \mathbb{C}P^k$, denoted  $\mathcal{H}_{n,k}(\Sigma )$. There is a natural Riemannian metric on $\mathcal{H}_{n,k}(\Sigma )$ defined by the metrics $g$ and $h$ on $\Sigma $ and $\mathbb{C}P^k$ as

\begin{equation}
\gamma _{L^2} (X,Y)=\int_{\Sigma } h (X,Y)\text{vol}_{g},
\end{equation}

\noindent for $X,Y\in T_\phi\mathcal{H}_{n,k}(\Sigma )\subset \Gamma (\phi ^*T\mathbb{C}P^k)$. This is called the $L^2$ metric on  $\mathcal{H}_{n,k}(\Sigma )$.\\

\noindent In the physics literature, the  degree $n$  holomorphic map  $\phi$  is regarded as a $\mathbb{C}P^k$ lump of charge $n$ on $\Sigma $, that is, a degree $n$  minimal energy static solution of the field equations of the $\mathbb{C}P^k $ model on $\Sigma $. Hence,  the degree $n$ moduli space $\mathcal{M}_n$ of the $\mathbb{C}P^k$ model on $\Sigma $ is $\mathcal{H}_{n,k}(\Sigma )$.  The low energy dynamics of $\mathbb{C}P^k$ lumps is conjecturally approximated by geodesic motion on $\mathcal{M}_n$  with respect to the $L^2$ metric $\gamma _{L^2}$ \cite{Leese, Speight0,Ward}. A precise version of this conjecture is   proved for $\Sigma =T^2$ and $n\geq 2$ by Speight in \cite{Speight5}. \\

\noindent With respect to the $L^2$ metric, Baptista \cite{Baptista3} has given  conjectural formulae for the volume and the Einstein-Hilbert action of  $\mathcal{H}_{n,k}(\Sigma )$, provided $\Sigma $ has genus $\text{g}\leq n/2$,

\begin{align}
\text{Vol}(\mathcal{H}_{n,k}(\Sigma ),\gamma _{L^2})&=\frac{(k+1)^{\text{g}}}{m!}\biggl(\pi \text{Vol}(\Sigma ,g)\biggr)^m,\label{baptis1}\\
H(\mathcal{H}_{n,k}(\Sigma ),\gamma _{L^2})&=\frac{2 \pi (k+1)^{\text{g}}[m-2\text{g}+1]}{(m-1)!}\biggl(\pi \text{Vol}(\Sigma, g)\biggr)^{m-1},\label{baptis2}
\end{align}

\noindent where $m=(k+1)(n+1-\text{g})+\text{g}-1$ and $\text{Vol}(\Sigma ,g)$ is the volume of $\Sigma $. This conjecture is based on a singular limit relating the $\mathbb{C}P^k$ model on $\Sigma $ with  a gauged sigma model whose fields take values in $\mathbb{C}^{k+1}$ \cite{Baptista3}. More precisely, a one parameter family of metrics on the $n$-vortex moduli space, which is a compact K\"{a}hler manifold, are conjectured to converge, in a certain limit, to the $L^2$ metric on $\mathcal{H}_{n,k}(\Sigma )$. Such convergence has recently been established rigorously by Lui \cite{Lui} in the sense of Cheeger-Gromov, that is on each open set in some locally finite open cover of $\mathcal{H}_{n,k}(\Sigma )$. This convergence does not directly imply Baptista's conjectured formulae for the volume and the Einstein-Hilbert action of $\mathcal{H}_{n,k}(\Sigma )$, however.\\

\noindent In the case $n=1$ and $\Sigma =S^2$, Speight \cite{Speight0,Speight4} has determined an explicit formula for the $L^2$ metric on $\mathcal{H}_{1,k}(S^2 )$, and then  computed the volume of $\mathcal{H}_{1,k}(S^2 )$ for $k\geq 2$ finding agreement with the conjectural formula (\ref{baptis1}). In this  paper, an explicit formula for the Ricci curvature tensor, and then the scalar curvature on $(\mathcal{H}_{1,k}(S^2 ),\gamma _{L^2})$ have been determined for $k\geq 2$, by exploiting the K\"{a}hler property of the $L^2$ metric. The Einstein-Hilbert action of $\mathcal{H}_{1,k}(S^2 )$ with respect to the $L^2$ metric is computed for $k\geq 2$ confirming the formula (\ref{baptis2}).\\

\section{Degree $1$ Holomorphic Maps $S^2\rightarrow \mathbb{C}P^k$}

\noindent This section  reviews the geometric structure of $\mathcal{H}_{1,k}(S^2)$ introduced  in \cite{Speight4}. Let $S^2$ be the $2$-sphere equipped with the standard round metric and  let  $\phi $ be a degree $1$ holomorphic map $S^2\rightarrow \mathbb{C}P^k$. Introducing homogeneous coordinates $[z_0,z_1]$ on $\mathbb{C}P^1\cong S^2$, then such  degree $1$ map has the form

\begin{equation}\label{phi}
\phi ( [z_0,z_1] )=[a_0 z_0+b_0 z_1,\dots,a_k z_0+b_k z_1],
\end{equation}

\noindent where $(a_0,\dots,a_k)$ and $(b_0,\dots,b_k)$ are linearly independent in $\mathbb{C}^{k+1}$. Since  the elements $(\xi  a_0,\xi  b_0,\dots, \xi  a_k,\xi  b_k)\in \mathbb{C}^{2k+2}$, where $\xi \in \mathbb{C}^\times$,   determine the same  holomorphic map $\phi $, then there is an open inclusion $\mathcal{H}_{1,k}(S^2)\hookrightarrow  \mathbb{C}P^{2k+1}$ which is used to equip  $\mathcal{H}_{1,k}(S^2)$ with a topology, differentiable and complex structures.\\

\noindent  The isometry groups $U(2)$ and $U(k+1)$ of $\mathbb{C}P^{1}$ and $\mathbb{C}P^{k}$ respectively build an isometric action of $G=U(k+1)\times U(2)$ on $\mathcal{H}_{1,k}(S^2)$, that is, $\phi \rightarrow \sigma _2 \circ \phi \circ \sigma _1^{-1}$ where $\sigma _1$ and $\sigma _2$ are isometries of $\mathbb{C}P^{1}$ and $\mathbb{C}P^{k}$. Generically, each orbit of $G$ on  $\mathcal{H}_{1,k}(S^2)$ is a real codimension $1$ submanifold of $\mathcal{H}_{1,k}(S^2)$ and has a unique element  $\phi _\mu$  given by

\begin{equation}
\phi _\mu ([z_0, z_1])=[\mu z_0,z_1,0,\dots,0],\qquad \mu > 1.
\end{equation}

\noindent An exceptional orbit of real codimension $3$ occurs when $\mu =1$. This action decomposes $\mathcal{H}_{1,k}(S^2)$ into a one parameter family of orbits parametrized by $\mu \in [1,\infty )$. For $\mu >1$, the isotropy group of the orbit $G_\mu $ of $\phi _\mu $ is

\begin{equation}
K =\biggl\{\biggl(\begin{pmatrix}
 e^{i\alpha }       & 0 & 0 \\
0 & e^{i\beta  } & 0  \\
0        &  0 & U
\end{pmatrix},
\begin{pmatrix}
e^{i(\alpha+\delta ) }        & 0  \\
0 &  e^{i(\beta +\delta ) }
\end{pmatrix}
\biggr): \alpha ,\beta ,\delta \in \mathbb{R}, U\in U(k-1)\biggr\}.
\end{equation}

\noindent By the Orbit-Stabilizer Theorem, each orbit $G_\mu $ is diffeomorphic to $G/K$.\\

\noindent Now, let $\mathfrak{g}$ and $\mathfrak{k}$ denote the Lie algebras of $G$ and $K$ respectively and $\langle , \rangle $ be the $Ad(G)$ invariant inner product on $\mathfrak{g}$,

\begin{equation}\label{inn}
\langle (M_1, m_1), (M_2, m_2) \rangle = -\frac{1}{2} (\text{tr} M_1 M_2 +\text{tr} m_1 m_2),
\end{equation}

\noindent where $M_i \in \mathfrak{u}(k+1)$ and  $m_i\in \mathfrak{u}(2)$. The tangent space of $\mathcal{H}_{1,k}(S^2)$ at $\phi _\mu $ is

\begin{equation}
V_{\mu }:=T_{\phi _\mu } \mathcal{H}_{1,k}(S^2) = \langle \frac{\partial }{\partial \mu } \rangle \oplus \mathfrak{p},
\end{equation}

\noindent where $\mathfrak{p}$ is the the orthogonal complement of $\mathfrak{k}$ in $\mathfrak{g}$ with respect to $\langle, \rangle$. The space $\mathfrak{p}$ can be decomposed into $Ad(K)$ invariant subspaces 

\begin{equation}\label{p}
\mathfrak{p}= \mathfrak{p}_0\oplus \mathfrak{p}_\mu \oplus \tilde{\mathfrak{p}}_\mu \oplus \hat{\mathfrak{p}}\oplus \check{\mathfrak{p}},
\end{equation}

\noindent where $\mathfrak{p}_0$ is a $1$ real-dimensional space, $\mathfrak{p}_\mu $, $\tilde{\mathfrak{p}}_\mu $ are $1$-complex dimensional subspaces depending on $\mu $, and $\hat{\mathfrak{p}}$ and $\check{\mathfrak{p}}$ are $(k-1)$ complex dimensional subspaces. The definitions of these subspaces are included in the Appendix. It was shown in \cite{Speight4} that

\begin{pro}\label{Pro1}
\noindent Let $\gamma $ be a $G$ invariant  K\"{a}hler metric on $\mathcal{H}_{1,k}(S^2)$. Then, for $k\geq 2$, $\gamma $ is uniquely determined by the one parameter family of symmetric bilinear forms $\gamma _\mu :V_\mu \times V_\mu \rightarrow \mathbb{R}$ given by

\begin{equation}\label{Ahol}
\gamma _\mu  =A_0(\mu ) d\mu ^2+8\mu ^2 A_0(\mu ) \langle ,\rangle_{\mathfrak{p}_0} + A_1(\mu )\; \langle,\rangle_{\mathfrak{p}_\mu }+ A_2(\mu )\langle,\rangle_{\tilde{\mathfrak{p}}_\mu }+ A_3(\mu )\; \langle,\rangle_{\hat{\mathfrak{p}}} + A_4(\mu )\; \langle,\rangle_{\check{\mathfrak{p}}},
\end{equation}

\noindent where $A_0,\dots, A_4 $ are smooth positive functions of $\mu $ defined by a single function $A(\mu )$ and a positive constant $B$ as follows

\begin{align}
A_0(\mu )&=\frac{1}{4\mu }\;A'(\mu ),\qquad\quad & A_1(\mu )&=A_2(\mu )=\frac{\mu ^2-1}{\mu ^2+1}\;A(\mu ),\nonumber\\
 A_3(\mu )&=B+\frac{A(\mu )}{2},& A_4(\mu )&=B-\frac{A(\mu )}{2},\label{AAhol}
\end{align}

\noindent and  $\langle ,\rangle_{\mathfrak{p}_i}$ denote  the  induced  inner products of $\langle , \rangle$  on the $Ad(K)$ invariant  subspaces, given in (\ref{p}). 

\end{pro}

\noindent  For the $L^2$ metric $\gamma _{L^2}$ on $\mathcal{H}_{1,k}(S^2)$, the function $A(\mu )$ and the constant $B$ are

\begin{equation}\label{ALhol}
A_{L^2}(\mu )=\frac{16 \pi }{c_1 c_2}\;\frac{\mu ^4-4 \mu ^2 \log \mu-1 }{(\mu ^2-1)^2},\qquad B_{L^2}=\frac{8 \pi }{c_1 c_2},
\end{equation}

\noindent where $c_1$ and $c_2$ are the constant  holomorphic sectional curvatures of $g$ and $h$ respectively.

\noindent Another $G$ invariant K\"{a}hler metric on $\mathcal{H}_{1,k}(S^2)$ is the induced metric defined by the inclusion $\mathcal{H}_{1,k}(S^2)\hookrightarrow \mathbb{C}P^{2k+1}$, where $\mathbb{C}P^{2k+1}$ is given the Fubini-Study metric (of constant holomorphic sectional curvature $c$, say). We call this the Fubini-Study metric on $\mathcal{H}_{1,k}(S^2)$, denoted $\gamma _{FS}$. It is determined by

\begin{equation}\label{AFShol}
A_{FS}(\mu )=\frac{4}{c}\;\frac{\mu ^2-1}{\mu ^2+1},\qquad B_{FS}=\frac{2}{c}.
\end{equation}

\noindent The volume form of a $G$ invariant K\"{a}hler metric $\gamma$, determined as in (\ref{Ahol}) by the function $A(\mu )$ and the constant $B$, on $\mathcal{H}_{1,k}(S^2)$ is

\begin{equation}\label{vhol}
\text{vol}_\gamma =V(\mu ) \: d\mu \wedge  \text{vol}_{G/K},
\end{equation}

\noindent where

\begin{equation}\label{volumefunction}
V(\mu )= \frac{1}{\sqrt{2}} \;A(\mu )^2 \biggl( B^2-\frac{A(\mu )^2}{4}\biggr)^{k-1} A'(\mu ),
\end{equation}

\noindent and $\text{vol}_{G/K}$ is the volume form of $G/K$  with respect to the inner product $\langle , \rangle$, defined in (\ref{inn}). It was shown that for $k\geq 2$, every $G$ invariant K\"{a}hler metric $\gamma $ on $\mathcal{H}_{1,k}(S^2)$ has finite volume\cite{Speight4}. In fact, if $\lim_{\mu \rightarrow \infty } A(\mu )=2 B$, this volume is 

\begin{equation}\label{HolVol}
\text{Vol}(\mathcal{H}_{1,k}(S^2),\gamma )= \sqrt{2} (2 B  )^{2k+1}\frac{(k-1)! k!}{(2k+1)! } \;\text{Vol}(G/K)=\frac{ (2 B \pi )^{2k+1}}{(2k+1)! },
\end{equation}

\noindent where $\text{Vol}(G/K)$ is the volume of $G/K$ with respect to $\langle , \rangle$.\\


\section{ Ricci Curvature Tensor }

\noindent With respect to any $G$ invariant K\"{a}hler metric $\gamma $, determined as in Proposition \ref{Pro1}, on $\mathcal{H}_{1,k}(S^2)$, we determine an explicit formula for the Ricci curvature tensor $\rho $ as follows

\begin{pro}\label{HolRicci}
Let $\gamma $ be a $G$ invariant K\"{a}hler metric on $\mathcal{H}_{1,k}(S^2)$, determined as in (\ref{Ahol}) by the function $A(\mu )$ and the constant $B$. Then, the Ricci curvature tensor $\rho $ on $(\mathcal{H}_{1,k}(S^2),\gamma )$ with $k\geq 2$ is uniquely determined by the one parameter family of symmetric bilinear forms $\rho _\mu :V_\mu \times V_\mu \rightarrow \mathbb{R}$,  given by

\begin{equation}\label{same}
\rho_\mu    =C_0(\mu ) d\mu ^2+8\mu ^2 C_0(\mu )\langle ,\rangle_{\mathfrak{p}_0} + C_1(\mu )\; \langle,\rangle_{\mathfrak{p}_\mu }+ C_2(\mu )\langle,\rangle_{\tilde{\mathfrak{p}}_\mu }+ C_3(\mu )\; \langle,\rangle_{\hat{\mathfrak{p}}} + C_4(\mu )\; \langle,\rangle_{\check{\mathfrak{p}}},
\end{equation}

\noindent where $C_0,\dots,C_4$ are smooth functions of $\mu $, determined as in (\ref{AAhol}), by the function $C(\mu )$ and the constant $D$ given by

\begin{equation}\label{Chol}
C(\mu )=4(k+1)\frac{\mu ^2-1}{\mu ^2+1}-2\mu  \frac{F'(\mu )}{F(\mu )},\quad\quad \quad D=2 (k+1),
\end{equation}

\noindent where

\begin{equation}
F(\mu )= \frac{A(\mu )^2  A'(\mu )}{A_{FS}(\mu )^2  A_{FS}'(\mu )}\biggl( B^2-\frac{A(\mu )^2}{4}\biggr)^{k-1}\biggl( B_{FS}^2-\frac{A_{FS}(\mu )^2}{4}\biggr)^{-(k-1)}.
\end{equation}

\end{pro}

\paragraph{Proof:}

\noindent The Ricci curvature tensor $\rho $ on $(\mathcal{H}_{1,k}(S^2),\gamma )$ is a $G$ invariant symmetric $(0,2)$ tensor which is  Hermitian and its associated 2-form $\hat{\rho }=\rho (J . , . )$ is closed. Hence, $\rho $  has the same structure as $\gamma $, that is, it is uniquely determined by the one parameter family of symmetric bilinear forms $\rho _\mu :V_\mu \times V_\mu \rightarrow \mathbb{R}$,  given  as in (\ref{Ahol}). Since the coefficients $C_0(\mu ),\dots,C_4(\mu )$ are defined as in (\ref{AAhol}) by a single function $C(\mu )$ and a constant $D$, then we only need to determine $C(\mu )$ and $D$. By Proposition \ref{Pro1}, we have

\begin{equation}\label{CD}
C(\mu )=C_3(\mu )-C_4(\mu ),\qquad D=\frac{1}{2} [C_3(\mu )+C_4(\mu )].
\end{equation}

\noindent To compute  $C(\mu )$ and $D$, we need first an orthonormal basis for $\mathfrak{p}$  with respect to the inner product $\langle , \rangle_{\mathfrak{p}}$. We shall use the orthonormal basis $\{Y_i,\hat{Y}_j,\check{Y}_j:i=0,\dots,4, j=1,\dots,2k-2\}$ introduced in \cite {Speight4}. The structure of this basis is included in the Appendix. Hence, the functions $C_3(\mu )$ and $C_4(\mu )$ can be given, for example, by

\begin{align}
C_3(\mu )&=\rho_\mu  (\hat{Y} _1, \hat{Y} _1)=-\rho_\mu  (J\hat{Y} _2, \hat{Y} _1)=\hat{\rho }_\mu (\hat{Y} _1, \hat{Y} _2),\nonumber\\
C_4(\mu )&=\rho_\mu  (\check{Y} _1, \check{Y} _1)=-\rho_\mu  (J\check{Y} _2, \check{Y} _1)=\hat{\rho }_\mu (\check{Y} _1, \check{Y} _2).\label{C34}
\end{align}

\noindent Now,  the volume form, given in (\ref{vhol}), of any $G$ invariant K\"{a}hler metric $\gamma $ on $\mathcal{H}_{1,k}(S^2)$ can be written as

\begin{equation}\label{vFv}
\text{vol}_\gamma = F(\mu )\:  \text{vol}_{\gamma _{FS}},
\end{equation}

\noindent where 

\begin{equation}\label{Fhol}
F(\mu )= \frac{A(\mu )^2  A'(\mu )}{A_{FS}(\mu )^2  A_{FS}'(\mu )}\biggl( B^2-\frac{A(\mu )^2}{4}\biggr)^{k-1}\biggl( B_{FS}^2-\frac{A_{FS}(\mu )^2}{4}\biggr)^{-(k-1)}.
\end{equation}

\noindent Hence, the Ricci form $\hat{\rho }$ with respect to $\gamma $ is \cite{Besse1},

\begin{equation}\label{phol}
\hat{\rho }=\hat{\rho }_{FS}-i\partial \bar{\partial } f,\qquad f(\mu ):=\log F(\mu ),
\end{equation}

\noindent where $\hat{\rho }_{FS}$ is the Ricci form with respect to $\gamma _{FS}$,  $\partial :\Omega ^{(p,q)}\rightarrow \Omega ^{(p+1,q)}$, and $\bar{\partial} :\Omega ^{(p,q)}\rightarrow \Omega ^{(p,q+1)}$ are the partial exterior derivatives on  the space of $(p,q)$-forms  $\Omega ^{(p,q)}$ on $\mathcal{H}_{1,k}(S^2)$. Using (\ref{phol}) in (\ref{C34}), we have

\begin{align}
C(\mu )&=\hat{\rho_\mu }_{FS} (\hat{Y} _1, \hat{Y} _2) -\hat{\rho_\mu }_{FS} (\check{Y} _1, \check{Y} _2)-i\bigl[(\partial \bar{\partial }f)_\mu (\hat{Y} _1, \hat{Y} _2) -(\partial \bar{\partial }f)_\mu (\check{Y} _1, \check{Y} _2) \bigr],\nonumber\\
&=C_{FS}(\mu )-i\bigl[(\partial \bar{\partial }f)_\mu (\hat{Y} _1, \hat{Y} _2) -(\partial \bar{\partial }f)_\mu (\check{Y} _1, \check{Y} _2)\bigr],\label{C}
\end{align}

\noindent and

\begin{align}
2D &=\hat{\rho_\mu }_{FS} (\hat{Y} _1, \hat{Y} _2) +\hat{\rho_\mu }_{FS} (\check{Y} _1, \check{Y} _2)-i[\bigl(\partial \bar{\partial }f)(\mu )(\hat{Y} _1, \hat{Y} _2) +(\partial \bar{\partial }f)(\mu )(\check{Y} _1, \check{Y} _2) \bigr],\nonumber\\
&=2 D_{FS}-i\bigl[(\partial \bar{\partial }f)_\mu (\hat{Y} _1, \hat{Y} _2) +(\partial \bar{\partial }f)_\mu (\check{Y} _1, \check{Y} _2)\bigr].\label{D}
\end{align}

\noindent Since $(\mathcal{H}_{1,k}(S^2),\gamma _{FS})$  is a $(2k+1)$ complex dimensional K\"{a}hler-Einstein manifold, then \cite{Kobayashi}

\begin{equation}\label{pFS}
\hat{\rho }_{FS}=c\; (k+1)\; \omega _{FS},
\end{equation}

\noindent where $\omega_{FS}$ is the K\"{a}hler form of $\gamma _{FS}$. Hence, the function $C_{FS}(\mu )$ and the constant $D_{FS}$  are

\begin{equation}\label{CFS}
C_{FS}(\mu )=c (k+1) A_{FS}(\mu )= 4(k+1) \frac{\mu ^2-1}{\mu ^2+1}, \quad D_{FS}=c (k+1) B_{FS}=2 (k+1).
\end{equation}

\noindent It remains to compute $(\partial \bar{\partial }f)_\mu (\hat{Y} _1, \hat{Y} _2) $ and $(\partial \bar{\partial }f)_\mu (\check{Y} _1, \check{Y} _2)$. Let  $\xi _0=-Y_0/(2\sqrt{2} \;\mu )$, then  the Hermiticity of $\gamma $ implies that  $J\xi _0=- \partial /\partial \mu $, and so,

\begin{equation}
(J^* d\mu) (\xi _0)= d\mu (J \xi _0)=d\mu (-\frac{\partial }{\partial \mu })=-1,
\end{equation}

\noindent where $J^*$ is the induced almost complex structure on $V_\mu ^*$. This means that  $\eta _0=-J^* d\mu $ is the covector  of $\xi _0$, that is, $\eta _0(\xi _0)=1$. The exterior derivative of $f$ is 

\begin{equation}
df=\frac{1}{2}\; f'(\mu )\;[(d\mu +i\eta _0)+(d\mu -i\eta _0)]=\frac{1}{2}\; f'(\mu )\;[(d\mu -iJ^* d\mu )+(d\mu +iJ^* d\mu )].
\end{equation}

\noindent This implies that the $(1,0)$-part $\partial f$ and the $(0,1)$-part $\bar{\partial} f$ of the 1-form $df$ are

\begin{equation}
\partial  f=\frac{1}{2}\; f'(\mu )\;(d\mu +i\eta _0),\quad\quad \bar{\partial}  f=\frac{1}{2}\; f'(\mu )\;(d\mu -i\eta _0).
\end{equation}

\noindent Since $d=\partial +\bar{\partial }$ and $\bar{\partial} ^2=0$, then

\begin{equation}\label{df2}
\partial \bar{\partial }f=d\bar{\partial }f= -\frac{i}{2} f''(\mu )\;d\mu \wedge \eta _0 -\frac{i}{2} f'(\mu ) d\eta _0,
\end{equation}

\noindent where $d\eta _0$ is a 2-form on $\mathcal{H}_{1,k}(S^2)$  given for any vector fields $X,Y $ on $\mathcal{H}_{1,k}(S^2)$ by 

\begin{equation}\label{eta2}
d\eta _0(X,Y)=X[\eta _0(Y)]-Y[\eta _0(X)]-\eta _0([X,Y]).
\end{equation}

\noindent Let $\xi _1,\xi _2$ be the extension of $\hat{Y}_1$ and $\hat{Y}_2$ as Killing vector fields on $\mathcal{H}_{1,k}(S^2)$. Then,  from (\ref{df2}) and (\ref{eta2}), we have

\begin{equation}\label{partial}
(\partial \bar{\partial }f)_\mu (\hat{Y}_1,\hat{Y}_2)=\frac{i}{2} f'(\mu )\; \eta _0([\xi _1, \xi _2]\Bigl\lvert_{\phi =\phi _\mu}) .
\end{equation}

\noindent The Lie bracket of Killing vector fields on $\mathcal{H}_{1,k}(S^2)$ can be defined by the Lie algebra bracket $[\;,\;]_{\mathfrak{g}}$ of $\mathfrak{g}$ as follows \cite{Besse2}

\begin{equation}
[\xi _1,\xi _2]\Bigl\lvert_{\phi =\phi _\mu}=-P _{\mathfrak{p}}([\hat{Y}_1,\hat{Y}_2]_{\mathfrak{g}}),
\end{equation}

\noindent where $P _{\mathfrak{p}}$ is the projection of $\mathfrak{g}$ to $\mathfrak{p}$. Since

\begin{equation}
\hat{Y}_{1}= (-E_{13} +E_{31}, \boldsymbol{0}),\quad\quad  \hat{Y}_{2}=i(E_{13} +E_{31}, \boldsymbol{0}),
\end{equation}

\noindent as in the Appendix. Then, we have

\begin{align}
[\hat{Y}_1,\hat{Y}_2 ]_{\mathfrak{g}}&= -  2 i ( E_{13} E_{31} -E_{31} E_{13}, \boldsymbol{0}),\nonumber\\
&=- i ( 2 E_{11} -2 E_{33}, \boldsymbol{0}),\nonumber\\
&= -\frac{i}{2} ( 3 E_{11} +E_{22}-2 E_{33}, e_{11}-e_{22}) + \frac{i}{2} (  E_{11} -E_{22}, -e_{11}+e_{22}),\nonumber\\
&= - \frac{i}{2} ( 3 E_{11} +E_{22}-2 E_{33}, e_{11}-e_{22}) + \frac{1}{\sqrt{2}} Y_0,
\end{align}

\noindent  where $E_{\alpha \beta }$ and $e_{\alpha \beta }$ denote  $(k+1)\times (k+1)$ and $2\times 2$  matrices respectively whose element $(\alpha ,\beta )$ is $1$, and the others being zero. Since the element $i( 3 E_{11} +E_{22}-2 E_{33}, e_{11}-e_{22})/2\in  \mathfrak{k}$, then it  vanishes under $P _{\mathfrak{p}}$, and so

\begin{equation}\label{xi1xi2}
[\xi _1,\xi _2]\Bigl\lvert_{\phi =\phi _\mu}=-\frac{1}{\sqrt{2}} Y_0.
\end{equation}

\noindent Substituting  (\ref{xi1xi2}) in (\ref{partial}), we get

\begin{equation}\label{logF1}
(\partial \bar{\partial }f)_\mu (\hat{Y} _1, \hat{Y} _2)= i\mu f'(\mu ).
\end{equation}

\noindent Similarly, one can find that

\begin{equation}\label{logF2}
(\partial \bar{\partial }f)_\mu (\check{Y} _1, \check{Y} _2)=-i\mu f'(\mu ).
\end{equation}

\noindent Substituting (\ref{CFS}), (\ref{logF1}) and (\ref{logF2}) in (\ref{C}) and (\ref{D}), we obtain  the function $C(\mu )$ and the constant $D$ as in (\ref{Chol}).

\hfill $\Box$ 


\section{ Scalar Curvature}

\noindent An orthonormal basis for $(V_\mu ,\gamma _\mu )$ can be defined as follows \cite{Speight4},

\begin{align}
X&=\frac{1}{\sqrt{A_0(\mu )}}\; \frac{\partial }{\partial \mu },\quad\quad & 
X_0&=\frac{1}{\sqrt{8\mu ^2 A_0(\mu )}}\;Y_0,\nonumber\\
X_1&=\frac{Y_1-\mu Y_3}{\sqrt{(1+\mu ^2) A_1(\mu )}},\quad\quad &
X_2&=\frac{Y_2+\mu Y_4}{\sqrt{(1+\mu ^2) A_1(\mu )}},\nonumber\\
X_3&=\frac{-\mu Y_1+ Y_3}{\sqrt{(1+\mu ^2) A_1(\mu )}},\quad\quad & 
X_4&=\frac{\mu  Y_2+ Y_4}{\sqrt{(1+\mu ^2) A_1(\mu )}},\nonumber\\
\hat{X}_j&=\frac{1}{\sqrt{A_3(\mu )}}\;\hat{Y }_j,\quad\quad & \check{X}_j&=\frac{1}{\sqrt{A_4(\mu )}}\;\check{Y}_j,\quad j=1,\dots, 2k-2.\label{orthonormal}
\end{align}

\begin{pro}\label{Holscalar}
Let $\gamma $ be a $G$ invariant K\"{a}hler metric on $\mathcal{H}_{1,k}(S^2)$, determined as in (\ref{Ahol}) by the function $A(\mu )$ and the constant $B$. Then, the scalar curvature of $(\mathcal{H}_{1,k}(S^2),\gamma )$  for $k\geq 2$ is

\begin{equation}\label{khol}
\kappa (\mu )= 2\biggl[2\frac{C(\mu )}{A(\mu )}+\frac{C'(\mu )}{A'(\mu )} \biggr]+2(k-1) \biggl[\frac{4(k+1)+C(\mu )}{2B+A(\mu )}+\frac{4(k+1)-C(\mu )}{2B-A(\mu )}\biggr].
\end{equation}

\end{pro}

\paragraph{Proof:}

\noindent The scalar curvature of a $G$ invariant K\"{a}hler metric $\gamma $, determined as in (\ref{Ahol}), with respect to the orthonormal basis (\ref{orthonormal}) is

\begin{align}
\kappa (\mu )=&\rho _\mu (X,X)+\sum_{i=0}^4 \rho _\mu (X_i,X_i)+\sum_{j=1}^{2k-2}\bigl[\rho _\mu (\hat{X_j},\hat{X_j})+\rho _\mu (\check{X_j},\check{X_j})\bigr],\nonumber\\
=&\frac{1}{ A_0(\mu )} \rho_\mu  (\frac{\partial }{\partial \mu },\frac{\partial }{\partial \mu })+\frac{1}{ 8 \mu ^2 A_0(\mu )}\rho_\mu  (Y_0,Y_0)+\frac{1}{A_1(\mu )}\;\sum_{i=1}^4\rho_\mu  (Y _i,Y _i)\nonumber\\
&+\frac{1}{A_3(\mu )}\;\sum_{j=1}^{2k-2} \rho_\mu  (\hat{Y }_j, \hat{Y }_j)+\frac{1}{A_4(\mu )}\;\sum_{j=1}^{2k-2} \rho_\mu  (\check{Y}_j, \check{Y }_j).\label{same2}
\end{align}

\noindent Using (\ref{same}) in (\ref{same2}), we get 

\begin{equation}
\kappa(\mu )=2\;\frac{C_0(\mu )}{A_0(\mu )}+4\;\frac{C_1(\mu )}{A_1(\mu )}+2(k-1)\;\biggl[\frac{C_3(\mu )}{A_3(\mu )}+\frac{C_4(\mu )}{A_4(\mu )}\biggr].
\end{equation}

\noindent Using the relations between the functions $A_i(\mu )$ and $C_i(\mu )$  with $A(\mu )$ and $C(\mu )$ respectively, as in (\ref{AAhol}), we obtain that the scalar curvature of a $G$ invariant  K\"{a}hler metric $\gamma $  on $\mathcal{H}_{1,k}(S^2)$ has the formula (\ref{khol}).

\hfill $\Box$


\section{Einstein-Hilbert Action of $\mathcal{H}_{1,k}(S^2)$}

\noindent The Einstein-Hilbert action of a Riemannian manifold $(M,g)$ is defined by the integral

\begin{equation}
H(M,g)= \int_M \kappa \:  \text{vol}_g,
\end{equation}

\noindent where $\kappa $ and $\text{vol}_g$  are   the scalar curvature and the volume form respectively with respect to the Riemannian metric $g$ on $M$.\\

\begin{theo}\label{HolEH}
The Einstein-Hilbert action of $\mathcal{H}_{1,k}(S^2)$ with respect to the $L^2$ metric $\gamma _{L^2}$ is

\begin{equation}\label{HolEHaction}
H(\mathcal{H}_{1,k}(S^2),\gamma _{L^2})=\frac{2^{2k+2}\; \pi ^{2k+1} (k+1) B_{L^2}^{2k} }{(2k)!}, \qquad \forall  \; k\geq 2.
\end{equation}

\end{theo}

\paragraph{Proof:}

\noindent In this proof, and for the rest of the paper, we will desist from denoting $\mu $ dependence explicitly in the functions $A(\mu )$ and $C(\mu )$.\\
\noindent  The Einstein-Hilbert action of $\mathcal{H}_{1,k}(S^2)$ with respect to any $G$ invariant K\"{a}hler metric $\gamma$ is

\begin{align}
H(\mathcal{H}_{1,k}(S^2),\gamma )
&=\int_{\mathcal{H}_{1,k}(S^2)}\: \kappa (\mu )\:V(\mu )\: d\mu \wedge \text{vol}_{G/K},\nonumber\\
&=\text{Vol}(G/K)\;\int_1^\infty  \: \kappa (\mu )\:V(\mu )\: d\mu,
\end{align}

\noindent  The scalar curvature of $(\mathcal{H}_{1,k}(S^2),\gamma )$, given in (\ref{khol}), can be written as

\begin{equation}
\kappa (\mu )= \frac{2}{A A'} \bigl[2 C A'+A C' \bigr]+(k-1)\biggl(B^2-\frac{A^2}{4}\biggr)^{-1} \bigl[4(k+1)B-A C\bigr],
\end{equation}

\noindent and then, by (\ref{volumefunction}), we have

\begin{align}
\kappa (\mu )\;V(\mu )&= \frac{2}{\sqrt{2}} \bigl[2 A C A'+A^2 C' \bigr]\;\biggl(B^2-\frac{A^2}{4}\biggr)^{k-1}\;\nonumber\\
&\:\;+\frac{(k-1)}{\sqrt{2}} A^2 A'\; \bigl[4(k+1)B-A C\bigr]\;\biggl(B^2-\frac{A^2}{4}\biggr)^{k-2},\nonumber\\
&=\frac{2}{\sqrt{2}}\;\biggl(B^2-\frac{A^2}{4}\biggr)^{k-1} \frac{d}{d\mu } (A^2\;C)-\frac{(k-1)}{\sqrt{2}}\; C A^3  A'\;\biggl(B^2-\frac{A^2}{4}\biggr)^{k-2}\nonumber\\
&\;\;+\frac{4(k^2-1)B}{\sqrt{2}} A^2 A'\;\biggl(B^2-\frac{A^2}{4}\biggr)^{k-2}.
\end{align}

\noindent Since

\begin{equation}
\frac{d}{d\mu }\biggl[\biggl(B^2-\frac{A^2}{4}\biggr)^{k-1} \biggr] =-\frac{(k-1)}{2} A\;A' \biggl(B^2-\frac{A^2}{4}\biggr)^{k-2},
\end{equation}

\noindent then,

\begin{equation}
\kappa (\mu )\;V(\mu )=\frac{2}{\sqrt{2}}\;\frac{d}{d\mu } \biggl[ A^2\;C\;\biggl(B^2-\frac{A^2}{4}\biggr)^{k-1} \biggr] +2 \sqrt{2}(k^2-1)B A^2\;A'\;\biggl(B^2-\frac{A^2}{4}\biggr)^{k-2}.
\end{equation}

\noindent Hence, the Einstein-Hilbert Action $H(\mathcal{H}_{1,k}(S^2),\gamma )$ is

\begin{align}
H(\mathcal{H}_{1,k}(S^2),\gamma )=&\frac{2}{\sqrt{2}}\text{Vol}(G/K)\;\biggl[ A^2\;C\;\biggl(B^2-\frac{A^2}{4}\biggr)^{k-1} \biggr]_1^\infty \nonumber\\
 &+ 2 \sqrt{2} (k^2-1)B^{2k-3} \text{Vol}(G/K)\; \int_{A(1)}^{A(\infty)}  \:A^2\;\biggl(1-\frac{A^2}{4B}\biggr)^{k-2} \;dA.
\end{align}

\noindent For the $L^2$ metric on $\mathcal{H}_{1,k}(S^2)$, the following limits  follow from (\ref{ALhol}),

\begin{align}
\lim_{\mu \rightarrow 1}A_{L^2}&=0,\qquad & 
 \qquad \lim_{\mu \rightarrow \infty }A_{L^2}&=2 B_{L^2},\nonumber\\
\lim_{\mu \rightarrow 1}C_{L^2}&=0,\qquad &
 \qquad \lim_{\mu \rightarrow \infty }C_{L^2}&=4(k+1),
\end{align}

\noindent and so,

\begin{equation}
\lim_{\mu \rightarrow 1}\biggl[ A_{L^2}^2\;C_{L^2}\;\biggl(B_{L^2}^2-\frac{A_{L^2}^2}{4}\biggr)^{k-1} \biggr]=\lim_{\mu \rightarrow \infty }\biggl[ A_{L^2}^2\;C_{L^2}\;\biggl(B_{L^2}^2-\frac{A_{L^2}^2}{4}\biggr)^{k-1} \biggr]=0.
\end{equation}

\noindent Thus, the Einstein-Hilbert Action with respect to the $L^2$ metric $\gamma _{L^2}$ on $\mathcal{H}_{1,k}(S^2)$ is 

\begin{equation}
H(\mathcal{H}_{1,k}(S^2),\gamma _{L^2})=2 \sqrt{2} \;(k^2-1)B_{L^2}^{2k-3}\;\text{Vol}(G/K)\;\int_{A_{L^2}(1)}^{A_{L^2}(\infty)}  \:A_{L^2}^2\;\biggl(1-\frac{A_{L^2}^2}{4B_{L^2}}\biggr)^{k-2} \;dA_{L^2}.
\end{equation}

\noindent To compute the integral above, let $t=A_{L^2}/2B_{L^2}$, then

\begin{equation}\label{Shol}
H(\mathcal{H}_{1,k}(S^2),\gamma _{L^2})=2^4 \sqrt{2} \; (k^2-1)B_{L^2}^{2k}\; \text{Vol}(G/K)\;\int_0^1  \:t^2\;\bigl(1-t^2\bigr)^{k-2} \;dt.
\end{equation}

\noindent The integral in (\ref{Shol}) is finite for all $k\geq 2$. In fact

\begin{equation}\label{pap2}
\int_0^1\;t^2\;\bigl[1-t^2\bigr]^{k-2} \;dt =\frac{2^{2k-2} (k-2)!\;k!}{(2k)!},\qquad \forall \;k\geq 2.
\end{equation}

\noindent The volume of $G/K$ can be extracted from the formula of $\text{Vol}(\mathcal{H}_{1,k}(S^2),\gamma )$ in (\ref{HolVol}), that is,

\begin{equation}\label{pap1}
\text{Vol}(G/K)=\frac{1}{\sqrt{2} }\frac{ \pi ^{2k+1}}{(k-1)! \; k!}.
\end{equation}

\noindent Substituting (\ref{pap2}) and (\ref{pap1}) in (\ref{Shol}), we get

\begin{equation}
H(\mathcal{H}_{1,k}(S^2),\gamma _{L^2})=\frac{2^{2k+2}\; \pi ^{2k+1} (k+1) B_{L^2}^{2k} }{(2k)!}.
\end{equation}

\hfill $\Box$

\noindent By taking the holomorphic sectional curvatures $c_1=c_2=4$, then the constant $B_{L^2}=\pi /2$, and so the Einstein-Hilbert action of $\mathcal{H}_{1,k}(S^2)$ with respect to the $L^2$ metric is

\begin{equation}
H(\mathcal{H}_{1,k}(S^2),\gamma _{L^2})=\frac{2^{2}\; \pi ^{4k+1} (k+1)}{(2k)!},
\end{equation}

\noindent which confirms  Baptista's conjectured formula (\ref{baptis2}).


\section*{Acknowledgements}

I would like to thank my supervisor  Martin Speight for constructive suggestions and useful discussions. Also, I acknowledge King Abdulaziz University for a PhD scholarship in Pure Mathematics.


\section*{Appendix}

\noindent The orthogonal complement $\mathfrak{p}$ of the Lie algebra $\mathfrak{k}$ in $\mathfrak{g}$ decomposes into the $Ad(K)$ invariant subspaces \cite{Speight4}

\begin{equation}
\mathfrak{p}= \mathfrak{p}_0\oplus \mathfrak{p}_\mu \oplus \tilde{\mathfrak{p}}_\mu \oplus \hat{\mathfrak{p}}\oplus \check{\mathfrak{p}},
\end{equation}

\noindent where
\begin{align}
\mathfrak{p}_0&=\bigl \{ 
(\lambda \text{diag}(i ,-i,0,\dots,0, \text{diag}(-i, i) ): \lambda \in \mathbb{R} \bigr \}\equiv \mathbb{R},\\
\mathfrak{p}_\mu &=\biggl \{ \biggr(
\begin{pmatrix}
0        & x & 0 & \dots \\
-\bar{x} & 0 & 0 & \dots \\
0        &  0\\
\vdots   & \vdots 
\end{pmatrix},
\begin{pmatrix}
0        & \mu x  \\
-\mu \bar{x} & 0 
\end{pmatrix}
\biggr ): x \in \mathbb{C} \biggr \}\equiv \mathbb{C},\\
\tilde{\mathfrak{p}}_\mu &=\biggl \{ \biggr(
\begin{pmatrix}
0        & -\mu \bar{y} & 0 & \dots \\
\mu y & 0 & 0 & \dots \\
0        &  0\\
\vdots   & \vdots 
\end{pmatrix},
\begin{pmatrix}
0        & -\bar{y}  \\
y & 0 
\end{pmatrix}
\biggr ): y \in \mathbb{C} \biggr \}\equiv \mathbb{C},\\
\hat{\mathfrak{p}}&=\biggl \{ \biggr(
\begin{pmatrix}
0 & 0 &  & -{\bf{u}}^\dagger  \\
0 & 0 &  & \dots \\
{\bf{u}} & \vdots  \\ 
\end{pmatrix},
\boldsymbol{0}
\biggr ): {\bf{u}} \in \mathbb{C}^{k-1} \biggr \}\equiv \mathbb{C}^{k-1},\\
\check{\mathfrak{p}}&=\biggl \{ \biggr(
\begin{pmatrix}
0 & 0 &  &  \dots \\
0 & 0 &  & -{\bf{v}}^\dagger \\
\vdots & {\bf{v}}  \\ 
\end{pmatrix},
\boldsymbol{0}
\biggr ): {\bf{v}} \in \mathbb{C}^{k-1} \biggr \}\equiv \mathbb{C}^{k-1}.
\end{align}

\noindent The almost complex structure $J$ acts  on $\mathfrak{p}$ as

\begin{equation}
J:(\lambda ,x,y,{\bf u}, {\bf v})\mapsto 4 \mu  \lambda  \frac{\partial }{\partial \mu }+(0,ix, iy, i{\bf u}, i{\bf v}).
\end{equation}

\noindent An orthonormal basis for $\mathfrak{p}$ with respect to the inner product $\langle,\rangle_{\mathfrak{p}}$, defined by (\ref{inn}), is given as follows

\begin{align}
Y_0&=\frac{i}{\sqrt{2}}\;(E_{11}-E_{22}, -e_{11}+e_{22}),\nonumber\\
Y_1&=(E_{12}-E_{21}, \boldsymbol{0}),\quad &
Y_2&= i (E_{12}+E_{21}, \boldsymbol{0}),\nonumber\\
Y_3&=(\boldsymbol{0}, -e_{12}+e_{21}),\quad & 
Y_4&=i (\boldsymbol{0}, e_{12}+e_{21}),\nonumber\\
\hat{Y}_{2i-1}&= (-E_{1,i+2} +E_{i+2,1}, \boldsymbol{0}),\quad & 
\hat{Y}_{2i}&=i(E_{1,i+2} +E_{i+2,1}, \boldsymbol{0}),\;\;i=1,\dots,k-1\nonumber\\
\check{Y}_{2i-1}&= (-E_{2,i+2} +E_{i+2,2}, \boldsymbol{0}),\quad & 
\check{Y}_{2i}&=i(E_{2,i+2} +E_{i+2,2}, \boldsymbol{0}),\;\;i=1,\dots,k-1,\label{YYY}
\end{align}

\noindent  where $E_{\alpha \beta }$ and $e_{\alpha \beta }$ denote  $(k+1)\times (k+1)$ and $2\times 2$  matrices respectively whose element $(\alpha ,\beta )$ is $1$, and the others being zero.\\

\end{document}